\documentclass[9pt, conference, a4]{IEEEtran}
\IEEEoverridecommandlockouts                              
\usepackage{array}
\usepackage{graphicx}
\usepackage{subcaption}
\usepackage[english]{babel}
\usepackage{xcolor}
\usepackage{amsmath}
\usepackage{amssymb}
\usepackage{cite}
\usepackage{rotating}
\usepackage{bm}

\ifCLASSINFOpdf
\else
\fi
\usepackage{amsmath}
\usepackage{multirow}
\usepackage{hyperref}
\usepackage{pifont}% http://ctan.org/pkg/pifont
\newcommand{\cmark}{\ding{51}}%
\newcommand{\xmark}{\ding{55}}%
\usepackage{array}
\usepackage{graphicx}
\usepackage{subcaption}
\usepackage[english]{babel}
\usepackage{xcolor}
\usepackage{amssymb}
\usepackage{cite}
\usepackage{rotating}
\begin{document}
%\title{Image Translation for Bladder Tissue Classification in Unbalanced Multi-Domain Endoscopic Images}
%\title{Unpaired Image-to-image Translation for Semi-supervised
%Bladder Tissue Classification in \\
%Multi-Domain Endoscopic Images}
%\title{Semi-supervised Generative Adversarial Network for Bladder Tissue Classification in \\ Multi-Domain Endoscopic Images}
\title{Semi-supervised Bladder Tissue Classification in \\
Multi-Domain Endoscopic Images}
%\title{Semi-supervised Bladder Tissue Classification in Multi-Domain Endoscopic Images}
%\author{\IEEEauthorblockN{Jorge F Lazo}
%\IEEEauthorblockA{School of Electrical and\\Computer Engineering\\
%Georgia Institute of Technology\\
%Atlanta, Georgia 30332--0250\\
%Email: http://www.michaelshell.org/contact.html}
%\and
%\IEEEauthorblockN{Homer Simpson}
%\IEEEauthorblockA{Twentieth Century Fox\\
%\and
%\IEEEauthorblockN{James Kirk\\ and Montgomery Scott}
%\IEEEauthorblockA{Starfleet Academy\\
%San Francisco, California 96678--2391\\


\author{\IEEEauthorblockN{Jorge F. Lazo\IEEEauthorrefmark{1}\IEEEauthorrefmark{2},
Benoit Rosa\IEEEauthorrefmark{2},
Michele Catellani \IEEEauthorrefmark{3}, 
Matteo Fontana \IEEEauthorrefmark{4}, 
Francesco A. Mistretta \IEEEauthorrefmark{4}, \\
Gennaro Musi \IEEEauthorrefmark{4},
Ottavio de Cobelli \IEEEauthorrefmark{4},
Michel de Mathelin\IEEEauthorrefmark{2} and
Elena De Momi\IEEEauthorrefmark{1}\IEEEauthorrefmark{4}}
\IEEEauthorblockA{\IEEEauthorrefmark{1}Department of Electronics, Information and Bioengineering,
Politecnico di Milano, Milan, Italy}
\IEEEauthorblockA{\IEEEauthorrefmark{2}ICube, UMR 7357 CNRS, 
Université de Strasbourg, Strasbourg, France
}
\IEEEauthorblockA{\IEEEauthorrefmark{3} Hospital Papa Giovanni XXIII, Bergamo, Italy}
\IEEEauthorblockA{\IEEEauthorrefmark{4} European Istitute of Oncology (IEO), IRCCS, Milan, Italy}
}


\maketitle

\begin{abstract}
Objective: 
Accurate visual classification of bladder tissue during  Trans-Urethral Resection of Bladder Tumor (TURBT) procedures is essential to improve early cancer diagnosis and treatment. 
During TURBT interventions,  White Light Imaging (WLI) and Narrow Band Imaging (NBI) techniques are used for lesion detection. Each imaging technique provides diverse visual information that allows clinicians to identify and classify cancerous lesions.
Computer vision methods that use both imaging techniques could improve endoscopic diagnosis. 
We address the challenge of tissue classification when annotations are available only in one domain, in our case WLI, and the endoscopic images correspond to an unpaired dataset, i.e. there is no exact equivalent for every image in both NBI and WLI domains.
Method:
We propose a semi-surprised Generative Adversarial Network (GAN)-based method composed of three main components: a teacher network trained on the labeled WLI data; a cycle-consistency GAN to perform unpaired image-to-image translation, and a multi-input student network.  
To ensure the quality of the synthetic images generated by the proposed GAN we perform a detailed quantitative, and qualitative analysis with the help of specialists. 
Conclusion:
The overall average classification accuracy, precision, and recall obtained with the proposed method for tissue classification are 0.90, 0.88, and 0.89 respectively, while the same metrics obtained in the unlabeled domain (NBI) are 0.92, 0.64, and 0.94 respectively. The quality of the generated images is reliable enough to deceive specialists. 
Significance:
This study shows the potential of using semi-supervised GAN-based bladder tissue classification when annotations are limited in multi-domain data.
The dataset is available in 
\url{https://zenodo.org/record/7741476#.ZBQUK7TMJ6k}
%We will release the dataset upon publication. 
\end{abstract}
\begin{IEEEkeywords}
\footnotesize{
bladder cancer, semi-supervised learning, generative-adversarial networks,  image-to-image translation, tissue classification, multi-domain image classification
} 
\end{IEEEkeywords}

% \fi
%

\IEEEpeerreviewmaketitle

\section{Introduction}
\label{sec:introduction}
Urinary tract cancer comprises different types of lesions ranging from benign tumors to aggressive neoplasms with high mortality. 
This disease had 164,000 patients reported in 2021 and it is among the top 10 most common cancers worldwide~\cite{siegel2021cancer}.
%Bladder cancer (BC), also referred to as urothelial carcinoma originates on the inner surface of the bladder and it can be classified in Muscle Invasive Bladder Cancer (MIBC) and Non-Muscle Invasive Bladder Cancer (NMIBC). 
Muscle Invasive Bladder Cancer originates on the inner surface of the bladder and is more likely to metastasize than Non-Muscle Invasive Bladder Cancer (NMIBC)~\cite{sanli2017bladder}.
%In addition to the previous mentioned classes, Carcinoma In Situ (CIS) is an entity on itself conformed of cells that may become cancer and spread. CIS is difficult to diagnose given its similarity to non-suspicious bladder tissue, and it is usually confirmed only after histological analysis. 
The gold standard for Bladder Cancer (BC) diagnosis is cystoscopy. In case of finding abnormal tissue, patients should undergo  Trans-Urethral Resection of the Bladder Tumor (TURBT)~\cite{degeorge2017bladder}. This procedure consists of the insertion of an endoscope in the urinary tract and the removal of visible tumor lesions.  

The World Health Organization WHO has defined a stratification of urothelial carcinoma accordingly to their propensity of invasion and it can be generalized into two main classes: High-Grade Carcinoma  (HGC) and Low-Grade Carcinoma (LGC)~\cite{ball2005pathology}. 
Visual classification of BC is a challenging task. The shapes of lesions either high-grade or low-grade tumors are quite similar in some cases, and the visual difference between healthy tissue and non-tumor lesions is not trivial~\cite{sanli2017bladder}. 
In fact, definitive diagnosis, staging, and grading of cancer are only possible after histological analysis of the resected tissue~\cite{hall2007guideline}. 

\begin{figure}[tbp]
    \centering
    \includegraphics[width=0.5\textwidth]{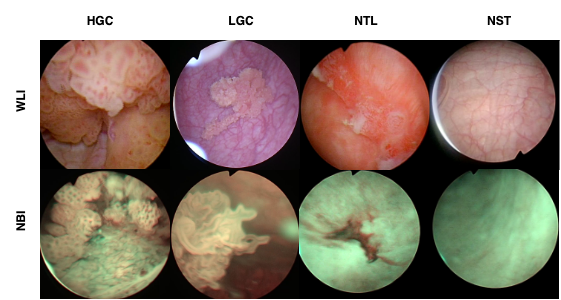}
    \caption{Sample images of the different classes in the bladder tissue classification dataset. From left to right: High-Grade Carcinoma (HGC), Low-Grade Carcinoma (LGC), No Tumor Lesion (NTL), and Non-Suspicious Tissue (NST).}
    \label{fig:tissue_samples}
\end{figure}

The use of different imaging techniques other than White Light Imaging (WLI), such as Narrow Band Imaging (NBI) can improve the differentiation of tumorous lesions from normal tissue~\cite{herr2014narrow, JEONG2018135}. 
Samples of different bladder tissue in both image domains are depicted in Fig.~\ref{fig:tissue_samples}. 
In NBI, a white light source is filtered in two narrow bands of 415 nm and 540 nm. At these wavelengths, the hemoglobin reflection spectra present a global and a local maximum respectively~\cite{hui2014wide}. 
This increases the contrast between the surface mucosa, the capillaries, and the blood vessels in the submucosa, therefore improving bladder cancer diagnosis by highlighting visual structures that are hard to notice when using only WLI~\cite{ye2015comparison}.  
Typically during TURBT procedures an initial inspection using WLI is carried out. 
Subsequently,  in a second inspection, the anatomical structures deemed suspicious are examined using NBI to confirm. 
In some cases, the use of NBI by itself could be more efficient than WLI in the detection of NMIBC~\cite{ye2015comparison}.

Despite the current advances in optical methods and their implementation in new devices, missing rates are reported to be between 10 and 20$\%$~\cite{chou2017comparative}.
The clinical interest in endoscopic tissue classification is related to the actions to be performed during surgery, as well as the follow-up treatment. 
The development of computer-aided diagnosis (CAD) systems for BC classification could help clinicians reduce current miss-classification rates which are related to incomplete excision of tumorous tissue, and cancer recurrence reported to have values of $75\%$~\cite{sylvester2006predicting}. 
For example, identifying a high-grade tumor in real-time could lead to the resection of a wider and deeper section of the tissue to avoid future recurrences. 

%This could be translated on better treatments given the extra information the operators could have available during the intervention. 
%For example, identifying a high grade tumor in real time could imply the resection of a wider and deeper section of the tissue to avoid future recurrences. 

In recent years, Deep-Learning (DL)-based methods have shown promising results in the analysis of endoscopic images. Most of the currently available datasets for endoscopic image analysis focus on colonoscopy~\cite{nogueira2021deep, KVASIR_dataset} and consist mainly of WLI data. 
Recently, few studies which include NBI data too have stressed on the advantage of using multi-domain data in the colonoscopy scenario~\cite{mesejo2016computer, KOMINAMI2016643,  xu2022deep}. 

In the case of the urinary system, only a few studies have been carried out in the task of tissue classification from endoscopic images~\cite{shkolyar2019augmented, yang2020automatic, ikeda2020support, ali2021deep}. Except for the study presented in~\cite{ali2021deep} where BL imaging is used, the rest of the studies use only WLI. 
%This impedes the exploitation of the different structural features that are more visible under different optical modalities, such as NBI, which could improve tissue classification methods. 
Multi-domain image classification implies several challenges, especially when the data and annotations are not evenly distributed across the different domains and some of the classes are under-represented~\cite{li2021multi}. 

In the specific case of TURBT some of these challenges include the fact that visually it is difficult to differentiate between lesions and the diagnosis is inconclusive~\cite{lingley2008computer}. 
Furthermore, due to the fact that multi-imaging endoscopes can collect only one imaging type at the time, it is not possible to have equivalent pairs of WLI and NBI images. 
Usually, an initial examination of the bladder is carried out using WLI and the lesions and anatomical structures deemed to be potentially cancerous tissue are examined again with NBI, in case this modality is available which is not always the case.  
An additional challenge is related to the imbalance of data in terms of the different classes and types of tissue. 
Non-Suspicious Tissue (NST) usually receives less attention during interventions, therefore fewer amount of image data is collected from it than from lesions,  either in WLI or NBI.  
Furthermore, non-cancerous lesions such as cystitis or other types of bladder inflammations are less common to appear in the initial inspection during TURBT.
All this contributes to the fact that most of the datasets (including ours) are imbalanced not only in terms of different image domains but also in terms of tissue classes. 

In this work, we focus on the task of bladder tissue classification in multi-domain images from TURBT procedures, with special emphasis in the fact that annotations only exist in one of these image domains. 
Considering that most state-of-the-art computer vision methods are sensitive to changes in domain~\cite{csurka2017comprehensive}, and the specific challenges existing in endoscopic image classification, we propose a GAN-based semi-supervised approach which comprises three main components: 
1)~A teacher network trained on the labeled WLI images. 
2)~A cycle consistency GAN to perform the unpaired image-to-image translation and 3)~A multi-input multi-domain image classifier trained in a semi-supervised way. We show that with our method it is possible to obtain satisfactory classification results even when annotations from one domain are not available.

%Unlike the studies presented in~\cite{zhao2019cycleemotiongan, chen2020cyclegan, hammami2020cycle, muramatsu2020improving, xu2019semi} where cycle-consistency translation has been implemented as a way of augmenting their datasets, we use image-translation inside a semi-supervised training loop to improve the classification performance of the unlabeled domain.  
To ensure that the images produced with the proposed translation network are consistent with the structural and pathological features of the source domain, we perform a detailed quantitative and qualitative analysis of the generative models. 
Additionally, we validate its quality with help of specialists familiar with the TURBT procedure.  
In order to allow future research in the task of bladder tissue classification, and ease benchmarking of future methods, we will release the dataset upon publication. 
%Moreover, since the endoscopic scenario requires that this translation is consistent with the structural and pathological features of the source domain, we propose the modification of the overall GAN loss by adding a structural similarity comparison which can be tuned to the specific source and target domain. 
%We perform a detailed quantitative and qualitative analysis of the images generated with different GAN models. 
%To quantitative select the overall best generative model we used the Fréchet Inception Distance (FID) and show that the semi-supervised classification performance improves with the proposed GAN method.  
%In a later stage, we show that this metric is in fact related to the tissue classification performance of the student network. 
%The qualitative evaluation was performed with the help of specialists familiar with TURBT procedure. 
%For this analysis we choose the GAN which obtained the best FID score and ask the participants to perform two different classification task using the generated images.  
%In order to allow future research in the task for bladder tissue classification, and ease benchmarking of future methods, we will release the dataset upon publication. 

%bladder tissue classification method which exploits the information of both image domains even when one of the domains is not available in the dataset. 
\section{Related Work}
\label{sec:related_work}

\subsection{Tissue Classification in Endoscopy}
\label{sec:sota_tissue_classification}

The analysis of endoscopic images has been rapidly developing in recent years thanks to the recent availability of new public datasets~\cite{misawa2021development, Pogorelov_2017}. 
In the specific task of tissue classification different models and techniques have been proposed with a special focus on the  gastrointestinal (GI) tract.
The existing methods range from the proposal of feature extraction models~\cite{pogorelov2017efficient, nadeem2018ensemble}, to the use of transfer learning and pre-trained CNNs~\cite{sanchez2020piccolo, ahmad2017endoscopic} and to more complex methods that focus on targeting the specific challenges present when working with GI endoscopic images~\cite{ali2020additive, struyvenberg2021computer, mohapatra2021wavelet, li2022adaptive}. 

In the case of the bladder, Ikeda et al.~\cite{ikeda2020support} proposed the use of 2-step transfer learning by first fine-tuning their models on 8728 gastroscopic images, and then re-training the models on 2102 cystoscopy WLI images, using the GoogLeNet model for the task of binary classification of images with and without NMIBC.
Yang et al.~\cite{yang2020automatic} compared the use of 3 different Convolutional Neural Networks (CNNs) as well as the platform EasyDL. The models used were LeNet, AlexNet and GoogLeNet. Their dataset  includes 1200 cystoscopy images with cancer and 1150 without. 
Shkolyar et al. ~\cite{shkolyar2019augmented} proposed CystoNet, a CNN for bladder cancer detection and binary classification. In their study, they used 2335 WLI frames of normal benign bladder mucosa and 417 histologically confirmed papillary urothelial carcinoma to train the network. 
In~\cite{lorencin2021urinary} the use of a Generative Adversarial Network (GAN) is proposed to perform data augmentation, then AlexNet and VGG16 are trained with the real and augmented data.  In total 202 images from a Confocal Laser Endomicroscope were used in their experiments.
In~\cite{ali2021deep} Ali et al. proposed the use of pre-trained models for the task of cancer malignancy, grading, and invasiveness classification on BL photodynamic cystoscopy images. The dataset was composed of 261 BL images and the pre-trained models used were VGG16, ResNet-50, MobileNetV2, and InceptionV3. On top of the pre-trained models, a shallow network was added to perform the classification.

\subsection{Image to Image Translation}
\label{sec:sota_im2im_translation}
Since its introduction, GANs have become an outstanding method for different tasks in DL applications. 
%Its fundamental principle is a zero-sum game where two networks compete against each other. The Generator $\mathcal{G}$ learns to generate realistic data while the other, the discriminator $\mathcal{D}$, should learn to differentiate between real and generated data.
GANs have been used for different purposes on endoscopic images such as the generation of synthetic images to improve polyp detection, or the construction of SLAM models to predict depth maps in colonoscopy~\cite{rau2019implicit, chen2019slam}.

One of the applications of GANs is image-to-image translation. This task can be resumed as the mapping of an image in domain $\mathbb{A}$ to another domain $\mathbb{B}$. In our case, these domains correspond to NBI and WLI. 
These types of models have been applied in diverse biomedical and endoscopic image tasks such as the translation between optical colonoscopy images and virtual colonoscopy images~\cite{mathew2020augmenting}, the mapping between cadaveric and live images~\cite{lin2020lc}, the adaptation between phantom images real endoscopic videos among others~\cite{sharan2021mutually, marzullo2021towards}. 

Using image-to-image translation with a focus on classification has been previously explored in other fields such as emotion classification, melanoma classification, and breast mass classification, among others.  
%This not only helps improving the classification goals of the network but also the quality of the images generated. 
In this regard, Yoo et al.~\cite{yoo2020joint} proposed a  joint learning approach using a mini-batch strategy and adaptive fade learning to use the generated images in the classifier with application in visually similar data. Likewise, Zhang et al.~\cite{zhang2021joint} and Mabu et al.~\cite{mabu2021semi} proposed the use of cycle consistency for classification in retinal pathologies identification and opacity classification in CT scans respectively. 

\subsection{Semi-Supervised Image Classification}
\label{sec:semi_supervised_class}
A common characteristic of medical image datasets is the lack of large annotated sets~\cite{cai2020review}. 
During the last few years semi-supervised learning methods have progressed as a good alternative to leverage this large amount of unlabeled data. 
One of the most common paradigms of semi-supervised learning is the use of Teacher-Student Networks~(TSN)~\cite{xie2020self}. In this type of model, a teacher network is trained on the labeled data, and a student network is trained on the unlabeled data using the predictions given by the teacher.
Training in semi-supervised mode allows the student model to learn features from unlabeled datasets~\cite{odena2016semi}.

In the endoscopic scenario, few studies have been carried out using semi-supervised learning. Du et al.~\cite{du2022improving} implemented a  semi-supervised contrastive learning method for Esophageal Disease Classification in a small dataset. Golhar et al.~\cite{golhar2020improving} proposed the use an unsupervised jigsaw learning method for GI lesion classification obtaining an improvement in accuracy of 9.8$\%$ with respect to supervised methods. 
Guo et al.~\cite{guo2020semi} proposed the use of a combination of a discriminative angular loss and Jensen-Shannon divergence loss for semi-supervised learning for wireless-capsule endoscopic image classification. Shi et al.~\cite{shi2021semi} implemented a TSN network for the 3D reconstruction of stereo endoscopic images. 

Recently, semi-supervised GAN-based models have been proposed for image classification in different fields such as natural images and hyper-spectral image classification~\cite{salimans2016improved, xue2020semi, li2020semi, wang2022ccs}.
However, in the field of endoscopic images it remains an unexplored topic. 

Unlike the studies presented in~\cite{zhao2019cycleemotiongan, chen2020cyclegan, hammami2020cycle, muramatsu2020improving, xu2019semi} where cycle-consistency translation has been implemented as a way of augmenting their datasets, we use image-translation inside a semi-supervised training loop to improve the classification performance of the unlabeled domain. 
Furthermore, the methods in which GAN-based semi-supervised methods have been proposed are mainly focused on the classification of images of the same domain. 
 
In this work, we propose a synergic semi-supervised GAN-based method that enables not only the exploitation of unlabeled data but also performs image translation to alleviate the dataset's domain imbalance. 
This allows the proposed network achieves a better generalization even in an image domain where labels are not available.   
%Further details of our implementation are explained in Sec.~\ref{sec:method}.
\section{Methods}
\label{sec:method}

Our overall goal is to improve tissue classification of endoscopic bladder images when labels are limited to only one domain, and there is no identical equivalent for every image on each domain. 
In our case, the endoscopic images are available on WLI and NBI domains, and the labels correspond only to the ones on WLI.
\begin{figure*}[tbp]
    \centering
    \includegraphics[width=0.99\textwidth]{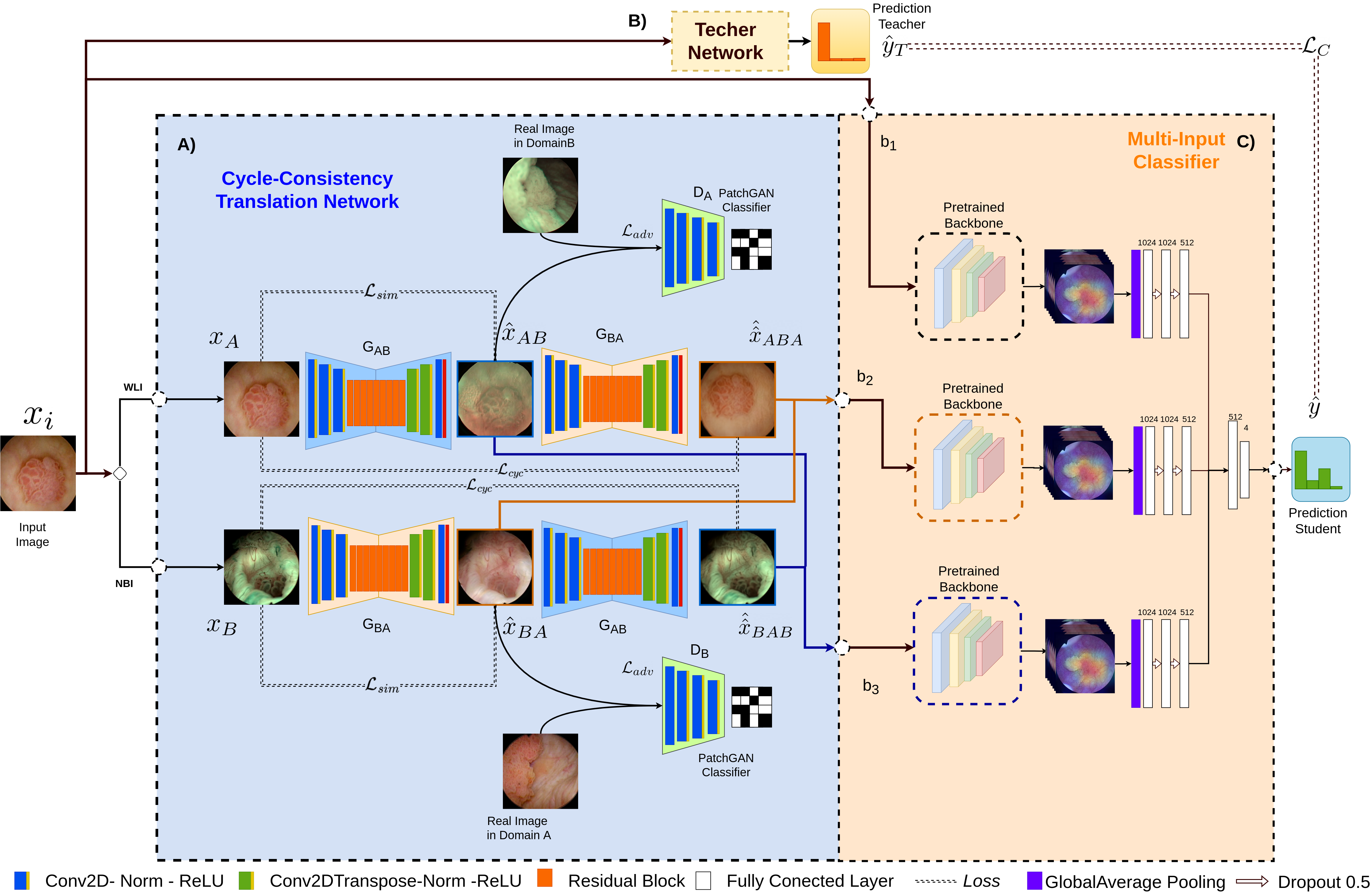}
    \caption{Proposed method. The network has two main elements. 
    A). Cycle-Consistency Translation Network that translates the image from NBI to WLI and vice-versa.
    B). Teacher network. 
    C). Multi-input network that performs the tissue classification task based on the features from both image modalities. The classification makes use of backbone networks that extract the features from each of the inputs to the classifier. The features are processed using Fully Connected (FC) layers which later are concatenated to perform the prediction in the final layer.}
    \label{fig:architecture}
\end{figure*}
\subsection{Problem Statement}

The proposed method consists of three main components; 1)~A cycle-consistency translation network to translate every image in the dataset and have equivalent paired images in both domains (NBI and WLI); 2)~A teacher network trained on the labeled WLI data; and 3)~A multi-input multi-domain classifier trained as student network in a TSN semi-supervised way. 
A schematic of the proposed model is depicted in Fig.~\ref{fig:architecture}.

Let us define a dataset $\mathcal{X} = \mathcal{X}_A \cup \mathcal{X}_B$ composed by the union of two subsets:  $\mathcal{X}_A=\{(\bm{x}_{A1}, \bm{y}_{A1}), ..., (\bm{x}_{An}, \bm{y}_{An})\}$ composed by $n$ labeled images $\bm{x}_i$ belonging to domain $\mathbb{A}$,  and $\mathcal{X}_B=\{\bm{x}_{B1}, \bm{x}_{B2}, ..., \bm{x}_{Bm}\}$ composed by $m$ unlabeled images $\bm{x}_j$ belonging to domain $\mathbb{B}$.
Initially, a classifier $\mathcal{C}$ is trained in a fully supervised fashion on $\mathcal{X}_A$. This classifier will work as a teacher model $\mathcal{C}_T$ at a later stage.
We propose the use of cycle-consistency image translation to deal with the issue of an unpaired and imbalanced dataset. For each image in domain $\bm{x}_A \in \mathbb{A}$ we will generate an equivalent translation $\bm{\hat{x}}_{AB} \in \mathbb{B}$, and for every $\bm{x}_B \in \mathbb{B}$ we will generate an equivalent translation $\bm{\hat{x}}_{BA} \in \mathbb{A}$. The translated images $\bm{\hat{x}}_{AB}$ and $\bm{\hat{x}}_{BA}$ are produced by the generators $\mathcal{G}_{AB}$ and $\mathcal{G}_{BA}$ respectively. 
An advantage of using cycle-consistency GANs is that an additional image $\bm{\hat{\hat{x}}}$ is generated, which corresponds to the reconstruction back to the original image. 
This can be used as additional data to train the student classifier. 
Therefore for every image $\bm{x}_{A}$ we have two extra images $\bm{\hat{x}}_{AB}$ and $\bm{\hat{\hat{x}}}_{ABA}$ and the same for  $\bm{x}_{B}$ where we have $\bm{\hat{x}}_{BA}$ and $\bm{\hat{\hat{x}}}_{BAB}$.
Then we train a multi-input classifier $\mathcal{C}_S$ which takes as input $\mathcal{C}_S(\bm{x}_{A}, \bm{\hat{x}}_{AB}, \bm{\hat{\hat{x}}}_{ABA})$ or $\mathcal{C}_S(\bm{x}_{B}, \bm{\hat{\hat{x}}}_{BA}, \bm{\hat{x}}_{BAB})$, depending on the domain of the input data. 
%This has the advantage that the network $\mathcal{C}_S$ is trained on the same amount of images from both domains, and consequently should have a better performance in both domains than a model that is trained on a single one. 

\subsection{Cycle-consistency Translation Network}

The unpaired image-to-image translation network is a generative adversarial network based on the \emph{Cycle}GAN architecture~\cite{zhu2017unpaired}. Two generators $\mathcal{G}_{AB}$ and $\mathcal{G}_{BA}$ are trained to learn the mappings between the domains $\mathbb{A}=$~WLI and $\mathbb{B}=$~NBI, such that $\mathcal{G}_{AB}: \mathbb{A}\rightarrow \mathbb{B}$ and $\mathcal{G}_{BA}: \mathbb{B}\rightarrow \mathbb{A}$. $\mathcal{D}_A$ and $\mathcal{D}_B$ are the two discriminators trained two distinguish between the real and fake images of each domain.
The proposed model uses three main losses, the adversarial loss $\mathcal{L}_{adv}$, the cycle consistency loss $\mathcal{L}_{cyc}$ and a similarity loss $\mathcal{L}_{sim}$.

The cycle loss $\mathcal{L}_{cyc}$ is defined as

\begin{equation}
        \mathcal{L}_{cyc}(\mathcal{G}_{pq}, \mathcal{G}_{pq}, \bm{x}_p) = \mathbb{E}_{x_p} ||\bm{x}_p -\mathcal{G}_{qp}(\mathcal{G}_{pq} (\bm{x}_p))||
\end{equation}
where the indexes $p, q$ represent the domain of the image and the domain to which is translated.
The adversarial loss for each generator $\mathcal{G}_{pq}$ and discriminator $\mathcal{D}_p$ is defined as
\begin{multline}
    \mathcal{L}_{adv}(\mathcal{G}_{pq}, \mathcal{D}_{p})
    = \mathbb{E}_{\hat{x}_p}
    [log(\mathcal{D}_{p}(\bm{\hat{x}}_p))] \\
    + \mathbb{E}_{x_p}[log(1-D_{p}( \mathcal{G}_{q}(\bm{x}_p)))] 
\end{multline}
To preserve the fine-grain details, such as the capillaries and inner blood vessels, that are related to the intrinsic pathology of each image domain and which are an essential visual cue for diagnosis assessment, we propose the addition to the cycle-consistency network a similarity loss $\mathcal{L}_{sim}$. This is defined as:
\begin{multline}
    \mathcal{L}_{sim}(\mathcal{G}_{AB}, \mathcal{G}_{BA}) = [1 -  \sum_i^N F(\bm{\hat{x}}_{Ai}, \mathcal{G}_{AB}(\bm{x}_{Ai}))] \\ 
    + [1 -  \sum_i^N F(\bm{\hat{x}}_{Bi}, \mathcal{G}_{BA}(\bm{x}_{Bi}))]
\end{multline}
where $\bm{x}_A \in \mathbb{A}$ and $\bm{x}_B \in \mathbb{B}$ correspond to the images form the $\mathbb{A}$ and $\mathbb{B}$ domains and the $ith$ refers index over the a set of images of $N$ elements. $\bm{\hat{x}}_A$ and $\bm{\hat{x}}_B$ correspond to the translated images by the generators. 
$F(\bm{x}, \bm{\hat{x}})$ is the structural similarity (SSIM) between images $x$ and $\hat{x}$ proposed in~\cite{wang2004image} as:
\begin{equation}
    F(\bm{x}, \bm{\hat{x}}) = \frac{(2\mu_x\mu_{\hat{x}} + c_1)(2\sigma_{x\hat{x}} + c_2)}{(\mu_x^2 + \mu_{\hat{x}}^2 + c_1)(\sigma_x^2 + \sigma_{\hat{x}}^2 + c_2)}
\end{equation}
Where $\sigma_{x, \hat{x}}$ is the covariance between $x$ and $\hat{x}$ :

\begin{equation}
    \sigma_{x, \hat{x}} = \frac{1}{m - 1} \sum(x_j - \mu_x)(\hat{x}_j - \mu_{\hat{x}})
\end{equation}
$m$ is the number of pixels; $x_j$ and $\hat{x}_j$ are the $j$th pixel of $\bm{x}$ and $\bm{\hat{x}}$ respectively; $\mu_x$, $\mu_{\hat{x}}$ and $\sigma_{x}$ and $\sigma_{\hat{x}}$ are the mean intensities and standard deviations of  $\bm{x}$ and $\bm{\hat{x}}$, and $c_1$ and $c_2$ are stabilizing constants to avoid singularities when $\mu_x^2 + \mu_{\hat{x}}^2 \approx 0 $ and $\sigma_x^2 + \sigma_{\hat{x}}^2 \approx 0$ respectively.

The overall objective function of the generative network is then defined as 
\begin{multline}
    \mathcal{L}(\mathcal{G}_{AB}, \mathcal{G}_{BA}, \mathcal{D}_A, 
    \mathcal{D}_B) = \mathcal{L}_{adv}(\mathcal{G}_{AB}, \mathcal{D}_A) \\
    +\mathcal{L}_{adv}(\mathcal{G}_{BA}, \mathcal{D}_B) 
    + \lambda_1 \mathcal{L}_{sim}(\mathcal{G}_{AB}, \mathcal{G}_{BA})  \\ 
    +\lambda_2 \mathcal{L}_{sim}(\mathcal{G}_{AB}, \mathcal{G}_{BA}) + \lambda_3 \mathcal{L}_{cyc}(\mathcal{G}_{AB}, \mathcal{G}_{BA}, x_A)  \\
    +\lambda_4 \mathcal{L}_{cyc}(\mathcal{G}_{BA}, \mathcal{G}_{AB}, x_B)
    %+ \lambda_3 \mathcal{L}_{C}
\end{multline}
where $\lambda_i$ are the hyper-parameters that balance the impact of the losses. 
The generators are trained to minimize the overall function and the discriminators to maximize it. 
The proposed \emph{Cycle}GAN with Similarity loss is termed CSi-GAN in the remainder of this paper, and the case in which $\lambda_1$ = $\lambda_2$ = 0 it reverts to the classical \emph{Cycle}GAN. 

\subsection{Semi supervised classification}

Initially, the teacher model $C_T$ is trained on WLI images in a fully supervised way. 
This could be seen as disconnecting the branch that goes from the input image $\bm{x}$ to the Cycle-Consistency Translation Network in Fig.~\ref{fig:architecture}, and training the network to optimize eq.~\ref{eq:cce} substituting the $\bm{\hat{y}}_{Ti}$ pseudo-labels with the labels $\bm{y}_{Ai}$ from set $\mathcal{X}_A$.
Afterward, the student model $C_S$ is trained using the labeled and unlabeled data using the predictions $\bm{\hat{y}}_T$ obtained from the teacher. 
The student network corresponds to a multi-input classifier that takes 3 images as input $C_S(\bm{x}, \bm{\hat{x}}, \bm{\hat{\hat{x}}})$ as depicted in Fig.~\ref{fig:architecture}-(C). 
The first one $\bm{x}$ is the original image from either WLI ($\bm{x}_A$) or NBI ($\bm{x}_B$) domains, the other two images correspond to the ones generated by the generators $\mathcal{G}_{AB}$ and $\mathcal{G}_{BA}$ respectively. 
In the case of the branch that takes as input $\bm{x}$, random data augmentation operations are applied which include random crop, random rotation, and flipping. 
Backbone networks $b_1$, $b_2$, and $b_3$, are used to extract the features of each of the 3 input images. 
In our case, we used as backbone ResNet-101 trained on \emph{ImageNet}. 
The extracted features from each of the backbones are processed separately using a shallow network composed of 3 Fully Connected (FC) layers. The outputs from these layers are concatenated together, from which finally the class prediction is performed in the final layer.
The classifier was trained to optimize the categorical cross-entropy loss defined as:
\begin{equation}
     \mathcal{L}_{C}(\bm{\hat{y}}_{Ti}, \bm{\hat{y}_i}) = - \sum_{i}  \bm{\hat{y}}_{Ti} \cdot {log (\bm{\hat{y}_i})}
\label{eq:cce}
\end{equation}
where $\bm{\hat{y}}_i$ is the predicted output from the student model, $\bm{\hat{y}}_{Ti}$ is the corresponding pseudo-label provided by the teacher network, and $i$ refers to the index over the classes.

\subsection{Dataset}
\label{sec:dataset}

For this study, endoscopic videos from 23 patients undergoing TURBT were collected, as well as the respective histopathological analysis from the resected lesions. 
The matching between the visual data and the histological results was done with the aid of an expert surgeon. The matching was performed by analyzing frame-by-frame the videos. The sections of the bladder from which lesions were resected during the surgical intervention were then identified. 
To avoid ambiguities of having multiple lesions of multiple types, only the frames in which individual lesions appeared were used in the dataset.
This procedure was performed on all the WLI video clips as well as 3 patients with NBI video data.  
In total 4 classes were defined. 
Taking into consideration the general classification of BC as defined in~\cite{sanli2017bladder} by the WHO and the International Society of Urological Pathology (ISUP), two categories were considered for cancerous tissue: Low-Grade Cancer (LGC) and High-Grade Cancer (HGC).
Additionally, 2 extra categories were considered for No Tumor Lesion (NTL) which comprehends cystitis, caused by infections or other inflammatory agents, and Non-Suspicious Tissue (NST).  The detailed statistics of the dataset are shown in Table~\ref{tab:dataset}.

The videos were acquired at the European Institute of Oncology (IEO) at Milan, Italy. Each patient signed an informed consent document approved by the IEO and in accordance with the Helsinki Declaration. No personal data was recorded.

\begin{table}[h!]
    \centering
    \caption{\footnotesize{Composition of the dataset considering two light modalities; White Light Imaging (WLI) and Narrow Band Imaging (NBI).}}
        \resizebox{0.46\textwidth}{!}
        {\begin{tabular}{c|c|c|c|c}
        %& type of tissue & \multicolumn{2}{imaging type}{c| & total} \\ \hline
        \multirow{2}{*}{Tissue type} & \multirow{2}{*}{No. of patient cases} & \multicolumn{3}{c}{No. of images} \\   \cline{3-5}
               &       & WLI & NBI & Total \\ \hline \hline 
        HGC   &  8 & 386 & 64  & 469 \\
        LGC   &  9 & 454 & 145 & 647 \\
        NST   &  5 & 439 & 75  & 504 \\ 
        NTL   &  5 & 97  & 37  & 134 \\ \hline
        Total & 23* & 1433 & 321 & 1754 \\
        \multicolumn{5}{l}{\scriptsize{*The total number of patient cases does not correspond to the sum of the second}}\\
        \multicolumn{5}{l}{\scriptsize{column since some of the patients had more than one type of lesion.}}\\
    \end{tabular}}
    \label{tab:dataset}
\end{table}

To determine if the use of more data helps to achieve better generalization when training the GAN networks, we used additional data from the datasets presented in~\cite{mesejo2016computer, sanchez2020piccolo} which contains endoscopic images from colonoscopy in NBI and WLI domains, and~\cite{lazo2021transfer} which contains unlabeled data from TURBT as well in NBI and WLI domains.   

\subsection{Model Implementation}

The model was trained in three steps. First, the cycle consistency GAN was trained for 150 epochs with an initial learning rate of $2\mathrm{e}{-4}$ and batch size of 1. 
The $\lambda$ hyperparameters were set to $\lambda_1$=$\lambda_2$=2.0, and $\lambda_3$=$\lambda_4$=1.0
The second step consisted of training the teacher classifier using the labeled dataset $\mathcal{X}_A$.
Once the GAN model and the teacher networks were trained, the multi-input classifier was trained setting the initial learning rate at $1\mathrm{e}{-5}$ using a batch size of 32.
The models were implemented using Tensorflow 2.5 in Python 3.6 and deployed on an Nvidia GeForce GTX 1080 GPU.
The training of the classifiers was repeated 10 times for each of the different experiments carried out in this study. 

For performance benchmarking of the classifiers, a hold-out strategy was used, 4 patient cases randomly chosen were held as test dataset. 
The rest of the dataset was divided randomly in a 75/25 ratio for training/validation. 
In the case of the GAN models, only the train dataset used for supervised classification was used during the training of the different combinations described in Table~\ref{tab:gan_dataset}. 
For the semi-supervised training apart from  using the labeled WLI images and unlabeled NBI, all NBI cystoscopy images described in~\cite{lazo2021transfer} were added to the training dataset. The test dataset for the semi-supervised task remained the same as the one used to test the performance of the teacher model. 

\subsection{Evaluation protocol}

Each of the different modules that comprise the proposed method was evaluated separately, and the best components of each one were chosen.

In contrast with other DL models that are trained to minimize a loss function, GAN models are trained to converge to an equilibrium between the generator and the discriminator networks. 
For this reason, there is no objective loss function to train this type of model, and compare their performance objectively~\cite{salimans2016improved}. 
However, there are some quantitative techniques that have been proposed to assess the performance of GAN models~\cite{borji2019pros}.
%In this work we 
%The details for both evaluations are discussed in subsections~\ref{sec:method_quantitative}~and~\ref{sec:method_qualitative}.

%Instead, a suite of qualitative and quantitative techniques have been developed to assess the performance of a GAN model based on the quality and diversity of the generated synthetic images.

\subsubsection{Quantitative Evaluation of the Generators} \label{sec:method_quantitative}
Generator models are usually evaluated based on the quality of the images they generate. However, this type of evaluation might not fully show the performance of the models and might be subjective due to biases of the reviewer~\cite{borji2019pros}. In this regard, some authors have proposed the use of different metrics such as the Inception score, to quantitatively evaluate the quality of the generated images~\cite{salimans2016improved}. 
In our specific case, we have the limitation that the dataset does not correspond to natural images, such as the ones on \emph{ImageNet} dataset, and therefore we can not apply the Inception score directly. We use instead the Fréchet Inception Distance~(FID) proposed in~\cite{heusel2017gans}, to quantify the performance of each generator trained and defined as:
\begin{multline}
    \mathit{d}^2 \left( (\mathbf{m}, \mathbf{C}), (\mathbf{m_\omega}, \mathbf{C_\omega}) \right) = \lVert  
\mathbf{m} - \mathbf{m_\omega} \rVert ^{2}_{2} \\ + Tr \left(\mathbf{C} + \mathbf{C_\omega} -2 (\mathbf{C} \mathbf{C_\omega)}^{1/2} \right)
\label{eq:fid}
\end{multline}
were $\mathbf{m}$, $\mathbf{C}$ are the mean and covariance obtained from the last pooling layer of an Inception model using sample images produced by the generative model respectively, and $\mathbf{m_\omega}$, $\mathbf{C_\omega}$ are the corresponding ones using images from the original dataset. 

We also analyze how the amount of data affects the quality of the images and the classifiers' performance. For this purpose, we use 3 different combinations of datasets coming from 4 different sources. The datasets composition is shown in Table ~\ref{tab:gan_dataset}.

To measure the sensitivity of the models depending on the amount of data used, we analyze the sensitivity to noise for each of the generative models trained on the different datasets as proposed in~\cite{bashkirova2019adversarial}. 
We added zero-mean Gaussian noise  $N(0,\sigma)$ in a range of $\sigma = [0.025, 0.05, 0.075, 0.1, 0.2]$ to the translation result before reconstruction. We compute the Mean Square pixel Error (MSE) of the reconstructed image with respect to the original image $\bm{x}_i$ and calculate the sensitivity (SN) using the equation: 
\begin{equation}
    SN = \frac{1}{N} \sum_{i=1}^N MSE(\mathcal{G}_p(\mathcal{G}_q(x_i) + N(0,\sigma) - \bm{x}_i)
    \label{eq:mse_sensitivity}
\end{equation}
%where
We compared the sensitivity for each of the generators in the proposed Cycle Similarity network (CSi-GAN) and the baseline \emph{CycleGAN}.
%We compare the baseline \emph{CycleGAN} with the proposed Cycle Similarity network (CSi-GAN)  using each of the datasets. 
\begin{table}[tbp]
    \centering
    \caption{Dataset composition used for training the GAN models. $\mathbb{D}_1$ corresponds to our dataset described in Sec.~\ref{sec:dataset}. $\mathbb{D}_2$ corresponds to a dataset composed only of external sources. $\mathbb{D}_3$ corresponds to the union of all the previously mentioned datasets.}
    \resizebox{0.48\textwidth}{!}{
    \begin{tabular}{c|c|ccc}
    \multirow{2}{*}{Dataset type} & \multirow{2}{*}{composition}  & \multicolumn{3}{c}{No. of images}\\ 
    & & NBI & WLI & Total \\ \hline 
        $\mathbb{D}_1$ & $\mathbb{D}_{A}$ & 1036 & 228 & 1264 \\
        $\mathbb{D}_2$ & $\mathbb{D}_{B} \cup \mathbb{D}_{C} \cup \mathbb{D}_{D}$ & 4592 & 2512 & 7104\\
        $\mathbb{D}_3$ & $\mathbb{D}_{A} \cup \mathbb{D}_{2}$ & 5628 & 2740 & 8368\\
    \multicolumn{2}{l}{\scriptsize{$\mathbb{D}_{A}$: our dataset. $\mathbb{D}_{B}$: dataset described on ~\cite{lazo2021transfer}. }}\\
    \multicolumn{2}{l}{\scriptsize{$\mathbb{D}_{C}$: dataset described on ~\cite{sanchez2020piccolo}. $\mathbb{D}_{D}$: dataset described on ~\cite{mesejo2016computer}}} \\
    \end{tabular}}
    \label{tab:gan_dataset}
\end{table}

\subsubsection{Evaluation by Medical Specialists}
\label{sec:method_qualitative}
Once the different GAN models were trained, the one with the best FID score was selected as the one to be used for human evaluation. 
With this analysis, we intended to confirm that the quality of the generated images is good enough to deceive experts, as well as to have a baseline to compare the classification performance of the models with respect to the ones from specialists. 

To qualitatively evaluate the utility of the images an online survey was set up where medical experts were asked to complete two tasks. 
In the first task, 20 pairs of randomly selected images were shown to the participants. 
Each image pair corresponded to two images from the same domain; one of them was an original image taken with the endoscope while the other corresponded to a translated image by the GAN. 
The participants were asked to identify which one was the original one, and which one was the generated one. For this task, NBI and WLI image pairs  were evenly distributed with 10 samples for each case. 
In the second task, 40 pairs of images were shown to the participants. 
The clinicians were asked to classify the images according to the 4 classes explained in section~\ref{sec:dataset}. 
Each image pair corresponded to one of the following options distributed in a 50/50 ratio: 
1) A pair of images that showed the same anatomical region at different times. In this case, the pair of images could correspond to two images of the same region and the same domain or two images of the same region but with a different domain, i.e. (NBI, NBI), (WLI, WLI) and (NBI, WLI). Each of the possible cases was evenly distributed. 
2) In the second option, again two images were shown that correspond to the same anatomical region at different times.
However, in this case one of the images was domain translated. The images used in this task were randomly chosen, taking into consideration having an even distribution of the 4 different tissue classes. Image pairs from options 1) and 2) were randomly ordered across the survey.

%An additional analysis to compare the attention maps of the images generated by the different GAN models using Grad-CAM~\cite{selvaraju2017grad} was as well performed in order to compare better the representations learned by each of the models. 

%\subsubsection{Effects of Domain Translation} 
%\subsubsection{Withdrawal of one Domain in the classification data}
\subsubsection{Evaluation of the Classifiers}
\label{sec:method_classification}
Once the GAN models were trained, we incorporate them into the general workflow using them as the base backbone to produce the multi-domain input images to feed the student classifier. 
The training was performed first in a fully supervised manner and then in a semi-supervised way using the previously trained teacher. 
To select the teacher model, diverse pre-trained models previously used in the literature were trained and the one with the best performance metrics was chosen as the teacher. %
We also performed ablation studies as well to demonstrate the utility of each of the elements of the proposed method. 
In the final stage, we train the multi-input classifier in a fully supervised way, using each of the previously trained generative models to determine whether there is a correlation between the classification performance and the quality of the generated images.

\begin{figure*}[tbp]
    \centering
    \includegraphics[width=0.99\textwidth]{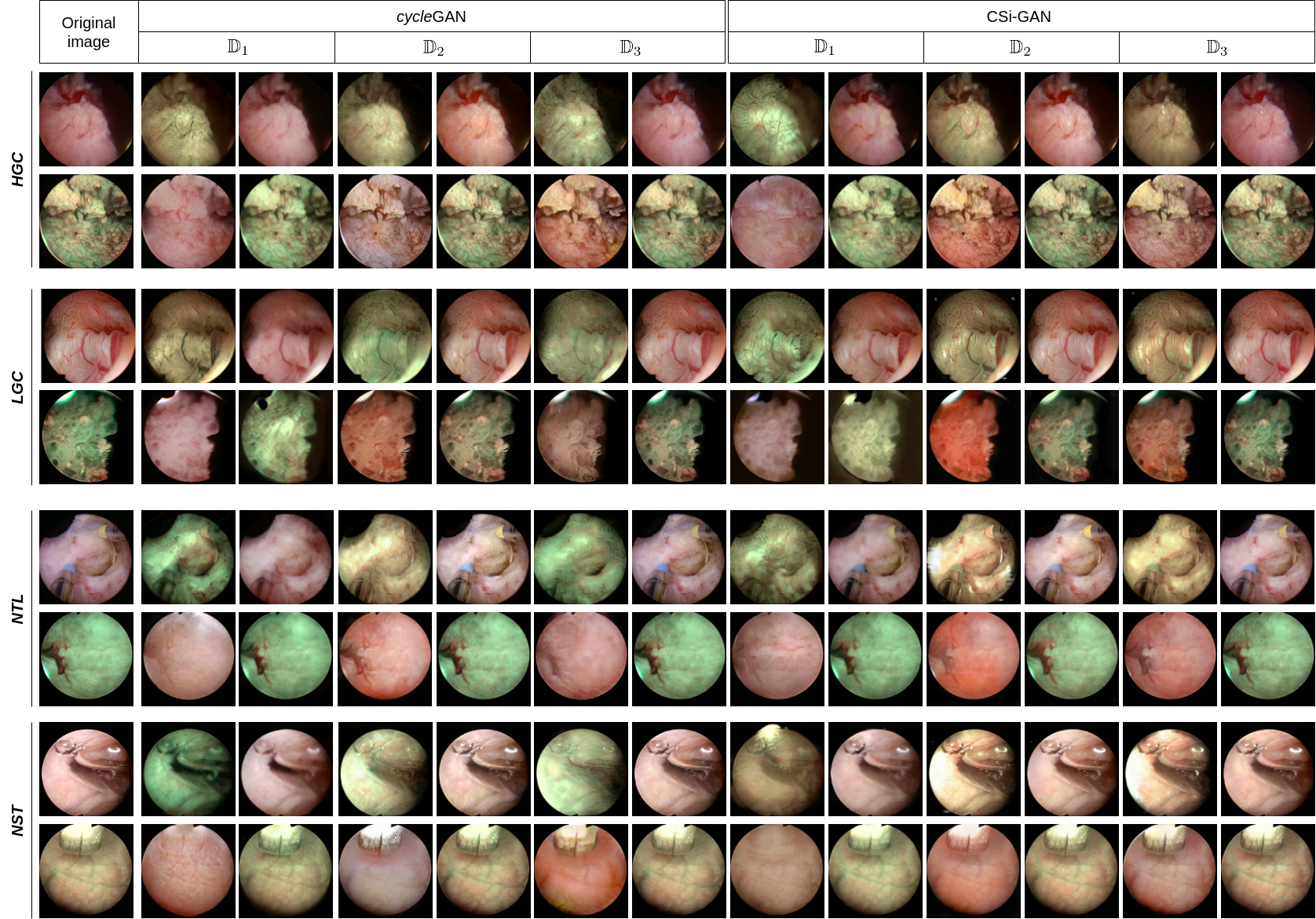}
    \caption{Samples of the generated images for the 4 classes on the 2 domains using each of the GAN models. 
    For each model trained on the 3 different datasets ($\mathbb{D}_1$, $\mathbb{D}_2$, $\mathbb{D}_3$) two images are shown: 1)~the translated image to the complementary domain, and 2)~the reversed translation back to its original domain. 
    }
    \label{fig:domain_translation_samples}
\end{figure*}

\subsection{Evaluation Metrics for Classification}

To evaluate the classification performance of the proposed method we used the metrics: accuracy ($Acc$), precision ($Prec$), recall ($Rec$), and F1-score. Additionally, as proposed in~\cite{guo2020semi} we also evaluated the model using Matthews correlation coefficient (MCC) and Cohen’s kappa (CK) statistic which has shown to be effective to benchmark diagnosis reliability of classifiers~\cite{saif2019abnormality}. 
Mann Whitney U-test was used to determine the statistical significance.
In the case of the user's experiments, the same metrics were used to evaluate their performance. 
Additionally, for the users' task of identifying the real images from the fake ones, the Area Under the Curve of the Receiver Operating Characteristic curve $AUC$ was used.

\section{Results and Discussion}
\label{sec:results}
This section is divided into two main subsections. First, we evaluate the performance of the image-translation network quantitative and qualitatively. Then we proceed to analyze the results of the classification network and the influence that the quality of the generated images has on the overall system, as well as the different components of the system.
\subsection{Evaluation of the GAN models}
\label{sec:qualitative_results}

The first set of results corresponds to the qualitative assessment of the synthetically generated images.
Samples of randomly chosen generated images by the different GAN models trained are shown in Fig.~\ref{fig:domain_translation_samples}.
A visual comparison shows that the amount and diversity of training data improve the quality of the images. 
We can observe that the addition of data helps the network learn the existence of other objects which do not correspond to the anatomical structures in the body, such as tools or bubbles. 
This shortcoming where the networks tend to disappear external structures by coloring them with the same hue as the rest of the background is more perceptible when models are trained with small datasets ($\mathbb{D}_{1}$). 
Furthermore, in these cases, the network also presents some noticeable flaws since sometimes the generated images present black dots scattered at diverse points. 
Nevertheless, the use of only external data ($\mathbb{D}_{2}$) also alters the hue of the translation.
This could be linked to the fact that the external data comes mainly from GI images which present different tints and anatomical formations than the ones present in the bladder.            
In general, for both cases \emph{cycle}GAN and CSi-GAN the use of the more general dataset ($\mathbb{D}_3$) which comprises data from the same anatomical target and external data produce the best quality images. However, still some image artifacts such as specularities, reflections, interlacing, etc. appear in the generated images without being present in the original one.  
The most significant improvement comes from using the $\mathcal{L}_{sim}$ loss to train the GANs. 
The fine-grain details, such as small vessels, are better preserved and highlighted after the translations, and it also helps to reduce the amount of noise in the image. 
%The same in case of some anatomical structures where the baseline GAN tend to fade its edges with the surrounding tissue. 
Similar behaviors can be observed in the video material attached to this manuscript. 
%This observations are confirmed by the quantitative results presented in section~\ref{sec:quantitative_results}.
%%%%%%%%%%%%%%%%%%%%%%%%%%%%%%%%%%%%%%%%%%%%%%%%%%%%%%%%%%%%%%%%%%%%%%%%%
\begin{table}[tbp]
    \centering
    \caption{FID scores and AUC of the Sensitivity curves for each of the GAN models trained on the different datasets. The results are divided in terms of the two generators $\mathcal{G}_{AB}$ and $\mathcal{G}_{BA}$. The numbers in bold indicate the cases that obtained the best metrics.}
    \begin{tabular}{cc|cc|cc}
        \multirow{2}{*}{model} & \multirow{2}{*}{dataset} & \multicolumn{2}{c}{FID} & \multicolumn{2}{c}{AUC} \\ 
        & & $\mathcal{G}_{AB}$ & $\mathcal{G}_{BA}$ & $\mathcal{G}_{AB}$ & $\mathcal{G}_{BA}$ \\\hline \hline
        \multirow{3}{*}{\emph{Cycle}GAN} & -$\mathbb{D}_1$ & 146.74 & 214.93 & 295.303 & 176.99\\
        & -$\mathbb{D}_2$ & 130.09 & 169.72 & 82.55 & 91.13 \\
        & -$\mathbb{D}_3$ & 138.79 & 164.24 & 113.69 & 119.57 \\ \hline
        \multirow{3}{*}{CSiGAN} & -$\mathbb{D}_1$ & 73.96 & 117.65 & 245.95 & 125.19 \\ 
        & -$\mathbb{D}_2$ & 54.33 & 72.13 & 81.76 & 80.18 \\
        & -$\mathbb{D}_3$ & \textbf{35.73} & \textbf{37.67} & \textbf{78.87} & \textbf{52.32}\\
    \end{tabular}
    \label{tab:FID_AUC_GANS}
\end{table}
\begin{table}[tbp]
    \centering
    \caption{Average results $\pm$ standard deviation from the specialist evaluation regarding their ability to discern between real and generated images. Results are divided in terms of the two different groups: Expert Surgeon (ES) and Resident (RE), and by the type of translation performed by each generator network i.e. WLI $\rightarrow$ NBI, NBI $\rightarrow$ WLI as well as the overall performance (ALL) of the GAN.}
    \resizebox{0.48\textwidth}{!}{\begin{tabular}{c|c|c|c|c|c}
        group ($n$) & Translation type & $Acc$ & $Prec$ & $Rec$  & $AUC$ \\ \hline
        \multirow{3}{*}{ES (15)}      
        & WLI $\rightarrow$ NBI & 0.66$\pm$0.13 & 0.66$\pm$0.18 & 0.8$\pm$0.20 & 0.59$\pm$0.14\\
        & NBI $\rightarrow$ WLI & 0.50$\pm$0.14 & 0.44$\pm$0.15 & 0.75$\pm$0.19 & 0.55$\pm$0.13 \\
        & ALL & 0.57$\pm$0.09   & 0.57$\pm$0.10 & 0.66$\pm$0.12 & 0.59$\pm$0.09 \\
        \multirow{3}{*}{RE (5)} 
        & WLI $\rightarrow$ NBI & 0.66$\pm$0.00 & 0.83$\pm$0.16 & 0.60$\pm$0.20 & 0.67$\pm$0.02 \\
        & NBI $\rightarrow$ WLI & 0.40$\pm$0.10 & 0.34$\pm$0.050 & 0.50$\pm$0.00 & 0.41$\pm$0.08 \\
        & ALL & 0.52$\pm$0.05 & 0.51$\pm$0.05 & 0.55$\pm$0.11 & 0.52$\pm$0.44 \\
    \end{tabular}}
    \label{tab:user_img_quality}
\end{table}
\subsubsection{Quantitative Evaluation of the GAN}
\label{sec:quantitative_results}
To evaluate the quality of the images generated by the GAN models the FID score and the AUC of the sensitivity curve were used. 
The results obtained for both metrics are shown in Table~\ref{tab:FID_AUC_GANS}. 
The model that obtains the best metrics for both cases, i.e. lower values, is the proposed CSi-GAN when trained on $\mathbb{D}_3$. 
In the case of FID score there is a clear difference between \emph{Cycle}GAN and CSi-GAN regardless of the dataset used for training,
with CSi-GAN obtaining in general better results.   
In the case of the AUC of the Sensitivity curve, the difference between the two models is not that obvious. This could be associated with the fact that neither of the networks is designed from the origin to be noise-resistant. 
However, there is a clear tendency that the addition of data makes CSiGAN more resistant to the addition of noise than its counterpart \emph{Cycle}GAN. 
This might be related to the fact that even if the addition of more data helps \emph{Cycle}GAN to generalize better in domain translation the lack of a structural loss inhibits it to discern properly between the correct information to produce a satisfactory translation, and the information that seems useful but is just noise. 
This could also explain the reason why \emph{Cycle}GAN obtains better metrics when trained on dataset $\mathbb{D}_2$ than on $\mathbb{D}_3$ since the quality of the images of $\mathbb{D}_2$ is higher and less noisy. 
\begin{figure}[h!]
    \centering
    \includegraphics[width=0.44\textwidth]{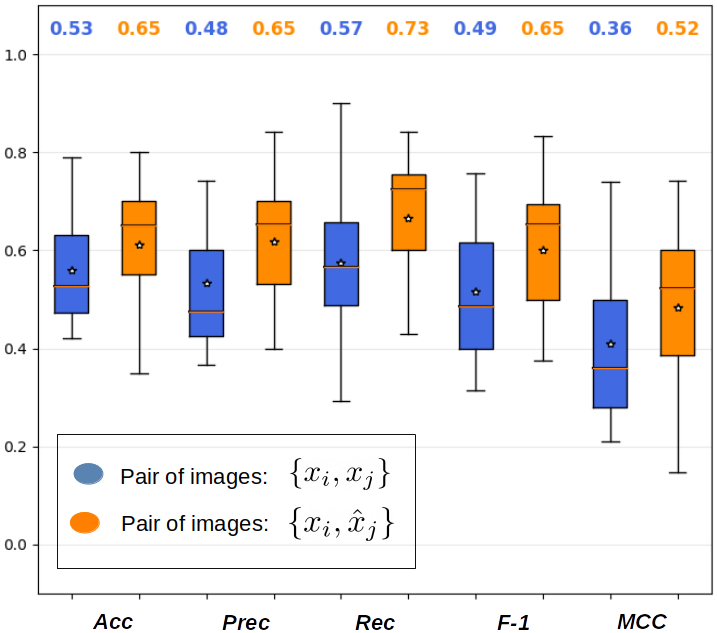}
    \caption{Box plot comparison of the surgeons performance in the tissue classification task. Blue boxes correspond to the case in which surgeons were shown a pair of real images $\{x_i, x_j\}$. Orange boxes correspond to cases in which a pair consisting of a real image $x$ and its translation $\hat{x}$ to the opposite domain $\{x_i, \hat{x}_j \}$, are shown.}    
    \label{fig:performance_users}
\end{figure}
\subsubsection{Evaluation by Medical Specialists}
\label{sec:medical_experts_eval}
In order to perform a more exhaustive analysis, a protocol was implemented to acquire feedback from expert clinicians in the field of endoscopy as described in sec.~\ref{sec:method_qualitative}. 
A total of 20 physicians from 10 different institutions familiar with TURBT participated in the study. 
Of this, 15 corresponded to Expert Surgeons (ES) and 5 to Residents (RE).
%For this analysis we choose the model that obtained the best FID score to generate the synthetic images, which correspond to the CSi-GAN trained on dataset $\mathbb{D}_{3}$.
For this analysis we choose the generative model which obtained the best FID score and AUC values, i.e. CSi-GAN trained on dataset $\mathbb{D}_{3}$, to generate the synthetic images.

The results regarding the ability of surgeons to discern between real and synthetic images are shown in table~\ref{tab:user_img_quality}. 
The results are split in 3 categories to evaluate separately each translation (WLI $\rightarrow$ NBI and  NBI $\rightarrow$ WLI) and therefore each generator independently, as well as the overall performance of the GAN (ALL).  
For both groups of participants (ES and RE), the results show slightly better results in the translation WLI $\rightarrow$ NBI for all metrics. This might be related to the fact that there are more sample images in the WLI training dataset than in the NBI and therefore the generator $\mathcal{G}_{AB}$ is able to generalize better and produce better quality images than its counterpart $\mathcal{G}_{BA}$.
The overall $AUC$ for ES is 0.59 and 0.52 for RE, meaning that their performance is marginally better than what a random binary classifier could achieve, confirming that the quality of the generated images is good enough to trick experts in the area. 

Concerning the tissue classification task, results are shown in Fig.~\ref{fig:performance_users}. 
%For reference we choose the best results of each group to compare with our method, the results are reported in table~\ref{tab:matrix_comparison}.
In the case of $Acc$ there was an average improvement of 8$\%$ when using a pair of a real image and a synthetic one than when only 2 real images were shown. In the case of $Prec$ the improvement was 19$\%$, while no improvement or decrease was observed in the case of $Rec$. For the F-1 score and $MCC$ the improvements were 16$\%$ and 17$\%$ respectively. 
However, no statistical significance was found. 
This goes in accordance with the results obtained in the previous analysis, meaning that the generated images do not affect the specialist's performance on tissue classification.

\begin{table*}[h!]
    \centering
    \caption{Comparison of using different pre-trained models in the proposed GAN-based multi-input classifier. The average $\pm$ standard deviation for each metric is presented in terms of the type of data in the test dataset (WLI and NBI) and the combination of both (ALL), for each of the models. The numbers in bold indicate the cases that obtained the best metrics.}
    \resizebox{\textwidth}{!}{
    \begin{tabular}{c|cccccccccc}
        \multirow{2}{*}{model} & \multirow{2}{*}{test data} & \multicolumn{3}{c}{$ACC$} & \multicolumn{3}{c}{F-1} & \multicolumn{3}{c}{$MCC$}  \\ 
        & & baseline & GAN-based & p-val & baseline & GAN-based & p-val& baseline & GAN-based & p-val\\ \hline \hline
        \multirow{3}{*}{VGG19} & NBI &  0.667$\pm$0.030 & 0.821$\pm$0.058 &           0.007 & 0.245$\pm$0.057 & 0.529$\pm$0.167 &      0.003 & 0.272$\pm$0.119 & 0.673$\pm$0.059 &      0.003\\
        & WLI & 0.653$\pm$0.048 & 0.667$\pm$0.033 &           0.789 & 0.675$\pm$0.034 & 0.716$\pm$0.057 &      0.060 & 0.487$\pm$0.054 & 0.564$\pm$0.084 &      0.298 \\
        & ALL & 0.661$\pm$0.033 & 0.684$\pm$0.031 &           0.286 & 0.649$\pm$0.022 & 0.649$\pm$0.052 &      0.797 & 0.567$\pm$0.038 & 0.572$\pm$0.053 &      0.298 \\ \hline
        \multirow{3}{*}{VGG16} & NBI & 0.692$\pm$0.056 & 0.744$\pm$0.075 &           0.018 & 0.409$\pm$0.144 & 0.409$\pm$0.174 &      0.325 &   0.010$\pm$0.212 & 0.376$\pm$0.237 &      0.060 \\
        & WLI &  0.720$\pm$0.022 &  0.740$\pm$0.025 &           0.014 & 0.641$\pm$0.046 & 0.716$\pm$0.025 &      0.002 &   0.610$\pm$0.030 & 0.632$\pm$0.035 &      0.006 \\
        & ALL & 0.714$\pm$0.017 & 0.741$\pm$0.028 &           0.002 & 0.648$\pm$0.046 & 0.741$\pm$0.023 &      0.001 & 0.602$\pm$0.024 & 0.634$\pm$0.036 &      0.001 \\ \hline
        \multirow{3}{*}{Inception V3} & NBI & 0.833$\pm$0.028 & 0.833$\pm$0.044 &           0.891 &  0.530$\pm$0.151 & 0.685$\pm$0.063 &      0.011 & 0.591$\pm$0.112 & 0.602$\pm$0.065 &    0.893 \\
        & WLI & 0.713$\pm$0.031 & 0.733$\pm$0.017 &           0.325 & 0.645$\pm$0.031 & 0.676$\pm$0.028 &      0.016 & 0.624$\pm$0.038 & 0.636$\pm$0.018 &      0.408 \\
        & ALL & 0.743$\pm$0.026 & 0.751$\pm$0.011 &           0.280 & 0.643$\pm$0.028 &  0.68$\pm$0.025 &      0.002 & 0.658$\pm$0.014 & 0.661$\pm$0.033 &       0.633 \\ \hline
        \multirow{3}{*}{Densenet} & NBI & 0.641$\pm$0.041 & 0.718$\pm$0.086 &           0.054 &  0.240$\pm$0.054 & 0.295$\pm$0.181 &      0.048 & 0.279$\pm$0.095 & 0.407$\pm$0.129 &      0.033 \\
        & WLI & 0.763$\pm$0.036 & 0.767$\pm$0.049 &           0.879 & 0.782$\pm$0.049 & 0.743$\pm$0.054 &      0.761 & 0.679$\pm$0.070 & 0.725$\pm$0.055 &      0.879 \\
        & ALL & 0.767$\pm$0.031 & 0.772$\pm$0.039 &           0.675 &  0.759$\pm$0.04 &  0.780$\pm$0.037 &      0.447 & 0.684$\pm$0.054 & 0.692$\pm$0.051 &     0.820 \\ \hline
        \multirow{3}{*}{ResNet-50} & NBI & 0.718$\pm$0.038 & 0.809$\pm$0.053 &           0.002 & 0.316$\pm$0.058 & 0.633$\pm$0.176 &      0.001 &  0.390$\pm$0.185 & 0.642$\pm$0.152 &      0.004 \\
        & WLI &  0.830$\pm$0.010 &  0.860$\pm$0.014 &           0.003 & 0.806$\pm$0.037 &  0.820$\pm$0.018 &      0.307 & 0.769$\pm$0.028 & 0.788$\pm$0.057 &     0.391 \\
        & ALL & 0.811$\pm$0.017 & 0.857$\pm$0.017 &           0.001 & 0.826$\pm$0.014 & 0.842$\pm$0.016 &      0.008 &  0.783$\pm$0.020 & 0.804$\pm$0.021 &      0.006 \\ \hline
        \multirow{3}{*}{ResNet-101} & NBI & 0.744$\pm$0.085 &  \textbf{0.862$\pm$0.046} &           0.011 & 0.452$\pm$0.242 & \textbf{0.713$\pm$0.174} &      0.016 & 0.547$\pm$0.196 & \textbf{0.757$\pm$0.081} &      0.008 \\
        & WLI & 0.861$\pm$0.027 &  \textbf{0.867$\pm$0.025} &           0.327 & 0.804$\pm$0.028 & \textbf{0.832$\pm$0.029} &      0.595 & 0.801$\pm$0.036 & \textbf{0.806$\pm$0.031} &      0.304 \\
        & ALL & 0.831$\pm$0.031 & \textbf{0.865$\pm$0.026} &           0.038 & 0.831$\pm$0.062 & \textbf{0.854$\pm$0.029} &      0.114 & 0.766$\pm$0.040 & \textbf{0.816$\pm$0.026} &      0.030 \\
    \end{tabular}}
    \label{tab:gans_vs_baseline}
\end{table*}

\subsection{Tissue Classification Evaluation}
\label{sec:results_tissue_classification}

Results regarding tissue classification are divided into three parts. 
First, we show that the use of our proposed GAN method for image translation improves in general the performance of tissue classification using different backbones previously used in the literature as simple fine-tuned classification networks. 
Next, we show that the use of semi-supervised learning, in general, improves further the classification performance.  
Finally, we perform an ablation analysis of the proposed model. 

\subsubsection{GAN-based Tissue Classification}
To test the generalization of our method, we compare the use of different networks (VGG16, VGG19, Inception V3, Desenet, ResNet-50, and ResNet-101) trained in a fine-tuning fashion against the implementation of these same networks in our GAN-based classification method.
%~\cite{ikeda2020support, yang2020automatic, shkolyar2019augmented}. 
CSi-GAN trained on $\mathbb{D}_3$ was chosen as the as the translation network. 
%Each of the networks  were used as backbones in the multi-input classifier network. 
%Results in terms of $ACC$, $MCC$ and F-1 score are shown in Fig.~\ref{fig:boxplots_base_acc_comparison}. 
Results in terms of $ACC$, $MCC$ and F-1 score are shown in Table~\ref{tab:gans_vs_baseline}.
Overall the use of the proposed GAN-based method obtains better metrics than the baseline networks.
%In all the metrics there is a significant improvement when using the CSi-GAN approach. 
In the majority of the cases, there is little improvement or no improvement when the input image is in the WLI domain. 
This uneven behavior in terms of the classification improvement might be related to the fact that WLI images are more similar to the natural images dataset in which the models were originally pre-trained (\emph{ImageNet)}. However, there is a noticeable improvement when it comes to the classification of NBI images where most of the base-line shows poor performances.  
%This improvement is on average of an increase of $15\%$ in the accuracy. The results are more noticeable when the backbones are $VGG19$ and $ResNet-50$. 
 
%\begin{figure*}[tbp]
%    \centering
%    \includegraphics[width=0.99\textwidth]{images/comparison_gan_vs_nogan.drawio.png}
%    \caption{Box plots comparisons in terms of $ACC$, F-1, $MCC$. Different base models were compared and implemented using the proposed CSiGAN .The models compared are A: \emph{Vgg19}, B: \emph{Vgg16}, C: \emph{Inception V3}, D: \emph{Densenet}, E: \emph{ResNet-50} F: \emph{ResNet-101}.
%    The results for each metric are divided in terms of the type of data in the test dataset (WLI and NBI) and the combination of both of them (ALL). The statistical significance using Mann Whitney U-test is denoted with $*:p<0.05$, $**:p<0.01$, $***:p<0.001$}
%    \label{fig:boxplots_base_acc_comparison}
%\end{figure*}

\subsubsection{Semi-supervised Classification}

\begin{figure*}[tbp]
    \centering
    \includegraphics[width=0.99\textwidth]{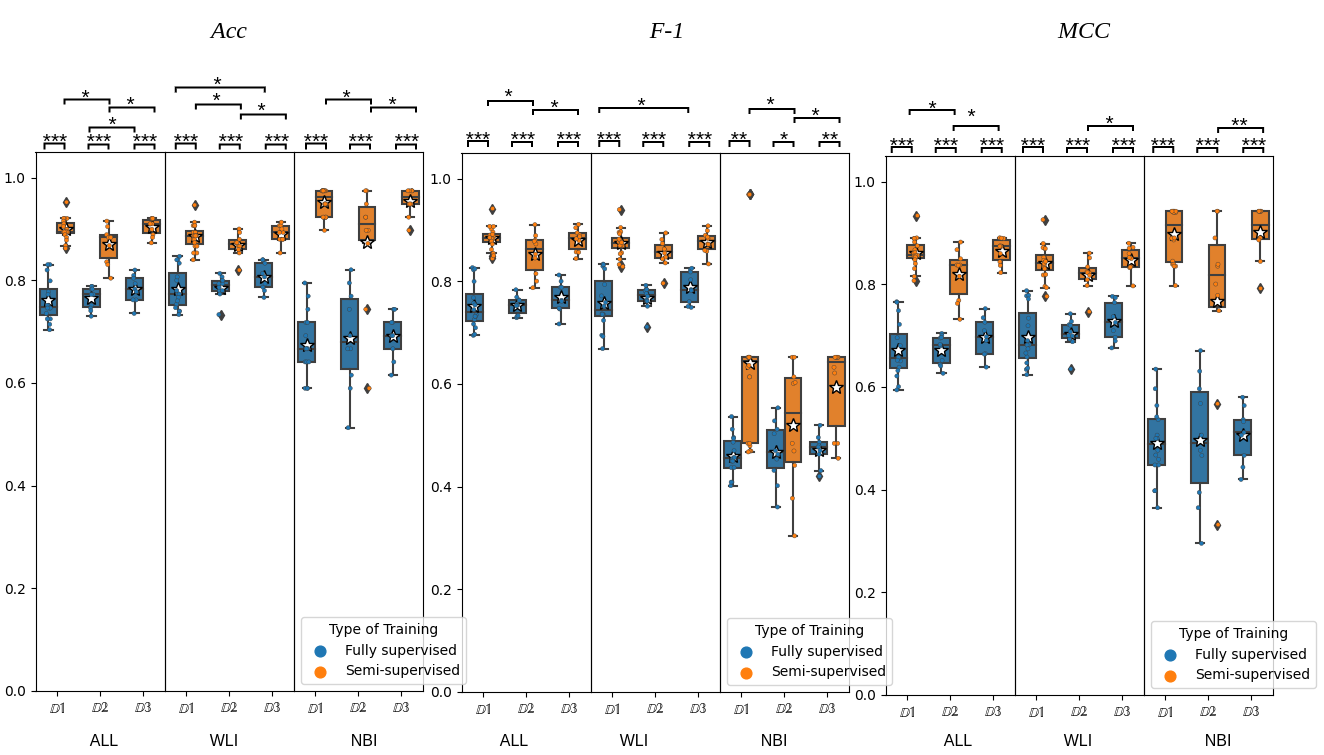}
    \caption{Boxplots comparison of $Acc$, $F-1$ score and $MCC$ of the proposed model trained in fully supervised vs semi-supervised way using CSi-GAN pre-trained on $\mathbb{D}_1$, $\mathbb{D}_2$ and $\mathbb{D}_3$.
    The results for each metric are divided in terms of the type of data in the test dataset (WLI and NBI) and the combination of both of them (ALL). The statistical significance using Mann Whitney U-test is denoted with $*:p<0.05$, $**:p<0.01$, $***:p<0.001$}
    \label{fig:semisup_vs_fullysup}
\end{figure*}

\begin{figure*}[htbp]
    \centering
    \includegraphics[width=0.99\textwidth]{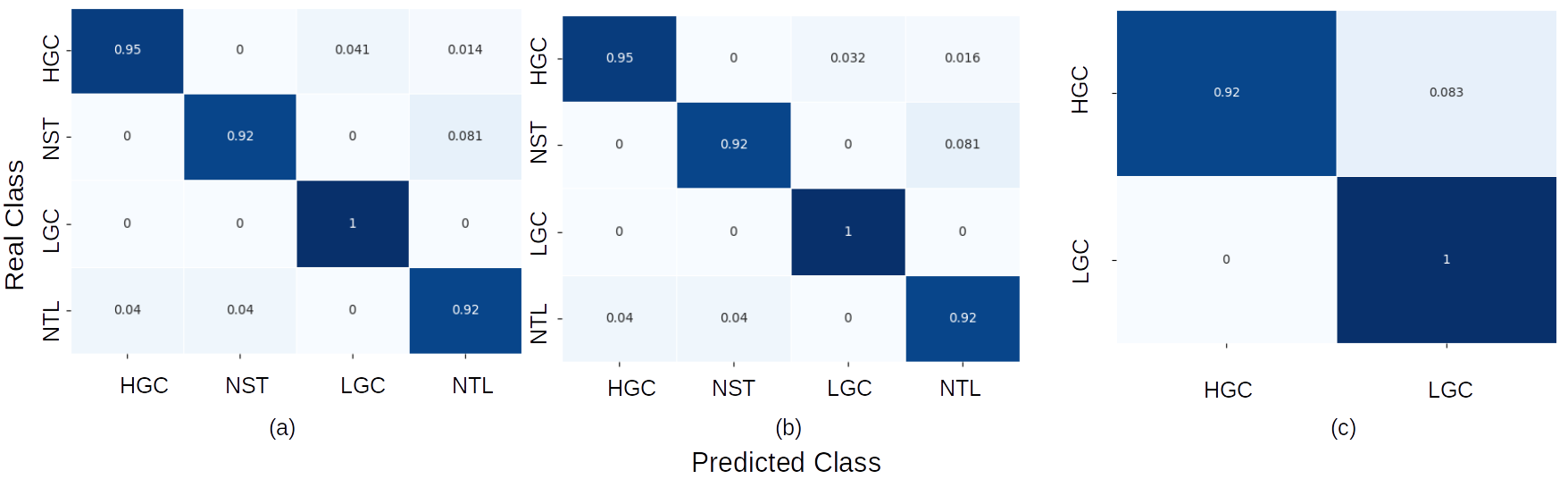}
    \caption{Confusion matrices of the best model obtained. a) Analysis on the complete test data (WLI + NBI). b) Analysis only on the WLI test data. c) Analysis on the NBI data. Is important to notice that due to the scarcity of annotated NBI data, the NBI test dataset was composed only of HGC and LHC images.}
    \label{fig:confusion_matrix_results}
\end{figure*}

We compared the use of GAN-based classification trained in a fully supervised way against the use of semi-supervised classification. 
In both cases, only the Multi-Input classifier weights were trained while the ones of the Cycle-Consistency Network remained constant. 
For these experiments, CSi-GAN pre-trained on each of the $\mathbb{D}_k$ datasets were used.
The results of these experiments are shown in Fig.~\ref{fig:semisup_vs_fullysup} in terms of $ACC$, F-1 score, and $MCC$. 
On average the improvement, in terms of $ACC$, F-1 score, and $MCC$, of using CSiGAN trained in a fully supervised way against the training in a semi-supervised fashion was of $8\%$, $6\%$, and $9\%$ respectively.
This shows the potential of using GAN-based semi-supervised learning for bladder tissue classification. 
The confusion matrices of the best model obtained are shown in Fig.~\ref{fig:confusion_matrix_results}.  

\begin{table*}[tbp]
    \centering
\caption{Ablation results. The average$\pm$ standard deviation for each metric is presented in terms of the type of data in the test dataset (WLI and NBI) and the combination of both (ALL), for each of the models. 
To have a reference point, the results obtained from physicians are shown too divided by specialists and residents. 
The table shows in which cases Domain Translation (DT) and Unlabeled Data (UD) were used during the training. 
The experiments to examine the impact of each of the branches ($b_1$m, $b_2$, $b_3$) in the multi-input classifier were performed in a fully supervised (FS) way in order to analyze the effects only of the translations performed by the GAN. 
The ablation results corresponding to branch $b_1$ is equivalent to the baseline (ResNet-101) result since the inputs from CSi-GAN are not used.   
The Cohen’s Kappa (CK) statistic is reported as an overall benchmark of the classifier. 
}
\resizebox{\textwidth}{!}{
\begin{tabular}{cccccccccccccccc}
method & UD & DT & test data &       Accuracy & p-val &      Precision & p-val  & Recall & p-val  &            F-1 & p-val &    $MCC$ & p-val &   CK & p-val  \\ \hline \hline
residents & - & - & ALL & 0.553$\pm$0.116 & - & 0.521$\pm$0.115 & - & 0.587$\pm$0.164 & - & 0.504$\pm$0.134 & - & 0.405$\pm$0.158 &  - & 0.385$\pm$0.157 &           - \\
specialist & - &  - & ALL & 0.579$\pm$0.111 & - & 0.542$\pm$0.113 & - &  0.607$\pm$0.16 & - & 0.523$\pm$0.132 & - & 0.424$\pm$0.153 & - & 0.418$\pm$0.151 & - \\
\multirow{3}{*}{baseline (ResNet-101)} & \multirow{3}{*}{\xmark} & \multirow{3}{*}{\xmark} & ALL & 0.831$\pm$0.031 & - & 0.843$\pm$0.019 & - & 0.831$\pm$0.062 & - & 0.831$\pm$0.031 & - &  0.766$\pm$0.04 & - & \multirow{3}{*}{0.762$\pm$0.044}  & \multirow{3}{*}{-} \\
&   &   &       WLI & 0.861$\pm$0.027 & - & 0.868$\pm$0.024 & - & 0.858$\pm$0.031 & - & 0.804$\pm$0.028 & - & 0.801$\pm$0.036 & - &   &  \\
&   &   &       NBI & 0.744$\pm$0.085 & - &  0.611$\pm$0.210 &  - &  0.85$\pm$0.095 &  - & 0.452$\pm$0.242 & - & 0.547$\pm$0.196 & - &  & \\
\multirow{3}{*}{CSi-GAN-$b_2$ (FS)} &       \multirow{3}{*}{\xmark} & \multirow{3}{*}{\cmark}&       ALL & 0.627$\pm$0.038 &          0.003 &  0.610$\pm$0.036 &           0.001 & 0.592$\pm$0.042 &        0.001 & 0.593$\pm$0.042 &     0.001 & 0.472$\pm$0.056 &             0.001 &  \multirow{3}{*}{0.47$\pm$0.057} &       \multirow{3}{*}{0.001} \\
              &         &                &       WLI &   0.610$\pm$0.030 &          0.003 &  0.587$\pm$0.030 &           0.001 & 0.572$\pm$0.032 &        0.001 & 0.565$\pm$0.034 &     0.001 & 0.455$\pm$0.042 &             0.001 &   &       \\
              &         &                &       NBI & 0.692$\pm$0.073 &            1.0 &  0.549$\pm$0.14 &           0.958 & 0.806$\pm$0.112 &        0.265 & 0.529$\pm$0.153 &     0.645 &  0.441$\pm$0.19 &             0.327 &   &       \\
\multirow{3}{*}{CSi-GAN-$b_3$ (FS)} &        \multirow{3}{*}{\xmark} & \multirow{3}{*}{\cmark} &        ALL & 0.688$\pm$0.026 &          0.001 & 0.706$\pm$0.024 &             0.001 & 0.691$\pm$0.026 &          0.001&   0.700$\pm$0.023 &       0.001& 0.563$\pm$0.037 &               0.001& \multirow{3}{*}{0.561$\pm$0.036} &         \multirow{3}{*}{0.001}\\
              &         &                &       WLI &   0.700$\pm$0.025 &          0.001 & 0.723$\pm$0.028 &             0.001& 0.723$\pm$0.019 &          0.001& 0.705$\pm$0.023 &       0.001&   0.610$\pm$0.030 &               0.001&  &          \\
              &         &                &       NBI & 0.641$\pm$0.058 &          0.114 & 0.487$\pm$0.112 &           0.287 &  0.840$\pm$0.084 &         0.61 & 0.404$\pm$0.067 &     0.391 & 0.483$\pm$0.069 &             0.298 & &          \\
\multirow{3}{*}{CSi-GAN (FS)} & \multirow{3}{*}{\xmark} & \multirow{3}{*}{\cmark} &       ALL &  0.865$\pm$0.020 &          0.038 & 0.849$\pm$0.017 & 0.210 & 0.853$\pm$0.0211 &        0.064 & 0.854$\pm$0.029 &      0.14 & 0.816$\pm$0.026 &              0.030 & \multirow{3}{*}{0.812$\pm$0.028} & \multirow{3}{*}{0.025} \\
&         &                &   WLI &  0.867$\pm$0.025 &          0.327 & 0.851$\pm$0.025 &           0.414 & 0.844$\pm$0.029 &        0.595 & 0.838$\pm$0.029 &     0.595 & 0.806$\pm$0.032 & 0.304 &  & \\
&         &                &       NBI & 0.872$\pm$0.046 &          0.011 & 0.839$\pm$0.023 &           0.771 & 0.921$\pm$0.054 &        0.137 &   0.713$\pm$0.174 &     0.016 & 0.757$\pm$0.081 &             0.008 &  & \\
\multirow{3}{*}{baseline semi-supervised} & \multirow{3}{*}{\cmark} & \multirow{3}{*}{\xmark} &       ALL & 0.868$\pm$0.019 &          0.018 & 0.853$\pm$0.024 &           0.077 &  0.856$\pm$0.02 &        0.059 & 0.849$\pm$0.021 &     0.028 & 0.817$\pm$0.026 &             0.024 & \multirow{3}{*}{0.815$\pm$0.026} & \multirow{3}{*}{0.017} \\
&          &               &       WLI & 0.863$\pm$0.015 &          0.731 & 0.864$\pm$0.016 &           0.926 &  0.841$\pm$0.021 &        0.239 & 0.847$\pm$0.017 &     0.476 & 0.809$\pm$0.021 &             0.598 &  &        \\
&          &               &       NBI & 0.803$\pm$0.075 &          0.027 & 0.615$\pm$0.146 &             1.0 & 0.848$\pm$0.058 &        0.082 &  0.614$\pm$0.16 &     0.456 & 0.835$\pm$0.154 &             0.072 &  &        \\
\multirow{3}{*}{SeCSi-GAN} & \multirow{3}{*}{\cmark} & \multirow{3}{*}{\cmark}  & ALL & \textbf{0.905$\pm$0.026} & 0.001 & \textbf{0.885$\pm$0.027} & 0.005 & \textbf{0.892$\pm$0.031} & 0.004 &  \textbf{0.889$\pm$0.031} &     0.002 & \textbf{0.867$\pm$0.036} &             0.001 & \multirow{3}{*}{\textbf{0.866$\pm$0.037}} & \multirow{3}{*}{0.001} \\
&          &                &       WLI & \textbf{0.897$\pm$0.016} &          0.001 & \textbf{0.887$\pm$0.019} &           0.012 & \textbf{0.895$\pm$0.022} &        0.005 &  \textbf{0.889$\pm$0.020} &     0.001 & \textbf{0.856$\pm$0.022} &             0.002 &  &     \\
&          &                &       NBI & \textbf{0.923$\pm$0.094} &           0.010 &   \textbf{0.640$\pm$0.093} &           0.075 &  \textbf{0.943$\pm$0.030} &        0.005 &  \textbf{0.762$\pm$0.160} &     0.086 &  \textbf{0.840$\pm$0.141} &             0.047 &  &      \\
\end{tabular}}
    \label{tab:domain_results}
\end{table*}

\begin{table*}[tbp]
    \caption{Ablation results in terms of each of the classes in the dataset.
    The average$\pm$ is the standard deviation of each metric for each of the 4 classes. 
    The experiments to examine the impact of each of the branches ($b_1$, $b_2$, $b_3$) in the multi-input classifier were performed in a fully supervised (FS) way in order to analyze the effects only of the translations performed by the GAN. 
    The ablation result corresponding to branch $b_1$ is equivalent to the baseline (ResNet-101) result since the inputs from CSi-GAN are not used.   }
    \centering
    \resizebox{0.80\textwidth}{!}{
    \begin{tabular}{cccccccccc}
    name model & metric &            HGC & p-val &            LGC & p-val &            NTL & p-val &            NST & p-val \\ \hline \hline 

    \multirow{3}{*}{baseline (ResNet-101)} &   $Prec$ &  0.86$\pm$0.068 &         - & 0.905$\pm$0.061 &         - &  0.683$\pm$0.09 &         - & \textbf{0.941$\pm$0.036} &         - \\
    &    $Rec$ & \textbf{0.919$\pm$0.078} &         - &  0.849$\pm$0.130 &         - &  0.869$\pm$0.084 &         - & 0.865$\pm$0.081 &         - \\
    &    F-1 & 0.854$\pm$0.044 &         - & 0.855$\pm$0.068 &         - & 0.761$\pm$0.054 &         - & 0.884$\pm$0.051 &         - \\
    \multirow{3}{*}{CSi-GAN-$b_2$}  &   $Prec$ &  0.630$\pm$0.068 &     0.003 & 0.598$\pm$0.048 &     0.001 & 0.487$\pm$0.066 &     0.013 & 0.709$\pm$0.035 &     0.003 \\
    &    $Rec$ & 0.669$\pm$0.064 &     0.005 & 0.708$\pm$0.099 &     0.151 &   0.300$\pm$0.082 &     0.003 &  0.770$\pm$0.059 &     0.254 \\
    &    F-1 & 0.647$\pm$0.059 &     0.001 & 0.628$\pm$0.048 &     0.001 &  0.367$\pm$0.08 &     0.003 & 0.736$\pm$0.029 &     0.003 \\
     
    \multirow{3}{*}{CSi-GAN-$b_3$} &   $Prec$ & 0.696$\pm$0.064 &     0.001 & 0.562$\pm$0.026 &       0.001&  0.630$\pm$0.109 &     0.247 & 0.912$\pm$0.037 &     0.176 \\
    &    $Rec$ &  0.649$\pm$0.060 &     0.002 &   0.660$\pm$0.050 &     0.032 &   0.560$\pm$0.060 &     0.002 & 0.865$\pm$0.013 &     0.731 \\
    &    F-1 & 0.671$\pm$0.047 &       0.001& 0.619$\pm$0.028 &       0.001& 0.605$\pm$0.069 &     0.001 & 0.877$\pm$0.014 &       1.0 \\
    
    \multirow{3}{*}{CSi-GAN}  &   $Prec$ & \textbf{0.919$\pm$0.029} &      0.020 & \textbf{0.943$\pm$0.036} &      0.260 & 0.606$\pm$0.077 &     0.125 & 0.925$\pm$0.041 &      0.410 \\
    &    $Rec$ & 0.885$\pm$0.031 &     0.319 &  0.868$\pm$0.070 &      0.230 &  0.880$\pm$0.081 &     0.972 & 0.824$\pm$0.062 &     0.723 \\
    &    F-1 & 0.901$\pm$0.018 &     0.056 & 0.912$\pm$0.037 &     0.044 & 0.704$\pm$0.043 &     0.125 & 0.863$\pm$0.037 &     0.864 \\

    \multirow{3}{*}{baseline semi-supervised} &   $Prec$ & 0.874$\pm$0.034 &     0.364 & 0.918$\pm$0.047 &     0.218 &  0.747$\pm$0.060 &     0.121 &  0.864$\pm$0.070 &     0.003 \\
    &    $Rec$ & \textbf{0.919$\pm$0.046} &     0.953 &  0.840$\pm$0.042 &     0.791 &  0.840$\pm$0.078 &     0.233 &  0.865$\pm$0.030 &      0.360 \\
    &    F-1 & 0.895$\pm$0.027 &     0.107 & 0.892$\pm$0.016 &     0.065 & 0.781$\pm$0.045 &     0.128 & 0.853$\pm$0.028 &     0.445 \\
    \multirow{3}{*}{SeCSi-GAN} &   $Prec$ & 0.914$\pm$0.053 &     0.013 & 0.926$\pm$0.058 &     0.814 &  \textbf{0.778$\pm$0.060} &     0.012 & \textbf{0.941$\pm$0.075} &     0.091 \\
    &    $Rec$ & \textbf{0.919$\pm$0.045} &     0.877 & \textbf{0.943$\pm$0.074} &     0.013 &    \textbf{0.880$\pm$0.015} &     0.072 &  \textbf{0.892$\pm$0.050} &     0.009 \\
    &    F-1 &  \textbf{0.914$\pm$0.040} &     0.001 & \textbf{0.922$\pm$0.044} &     0.002 &   \textbf{0.800$\pm$0.109} &     0.183 &  \textbf{0.895$\pm$0.040} &     0.409 \\

    \end{tabular}}
    \label{tab:results_per_class}
\end{table*}

\subsubsection{Ablation Results}
\label{sec:ablation_results}
In this case, we made a comparison between the base model, the proposed CSiGAN model trained in a fully supervised way, and in a semi-supervised way (Se-CSiGAN). 
We also analyzed the influence that each of the inputs of the multi-domain classifier model has. For this purpose, we trained the network with each of the individual branches ($b_1$, $b_2$, $b_3$) separately. 
%We performed this experiments in a fully-supervised way in order to isolate the effect of each branch  itself. 
The statistical significance was calculated with respect to the base-model ResNet-101.
Classification results obtained by medical experts, stratified between specialists and residents are shown as a reference point.  
%To have a reference point, the results obtained from medical experts are shown too divided by specialist and residents. 
The results of the ablation experiments are shown in the tables \ref{tab:domain_results} and \ref{tab:results_per_class}.
From these results, we can see that in general, all the models obtain better results than the specialists, and the major improvement comes from the use of a semi-supervised approach. 
However, the improvement obtained in the domain for which there are no labels when using domain translation is also noticeable.  
As expected, the integration of both results in the best performance, and improves considerably the detection of classes that are underrepresented. 
This behavior is more clearly noticeable in the case of the NTL class which in our dataset has the smallest number of samples and in contrast to NST could be easily misclassified as a tumorous lesion. 

\begin{figure*}
    \centering
    \includegraphics[width=0.85\textwidth]{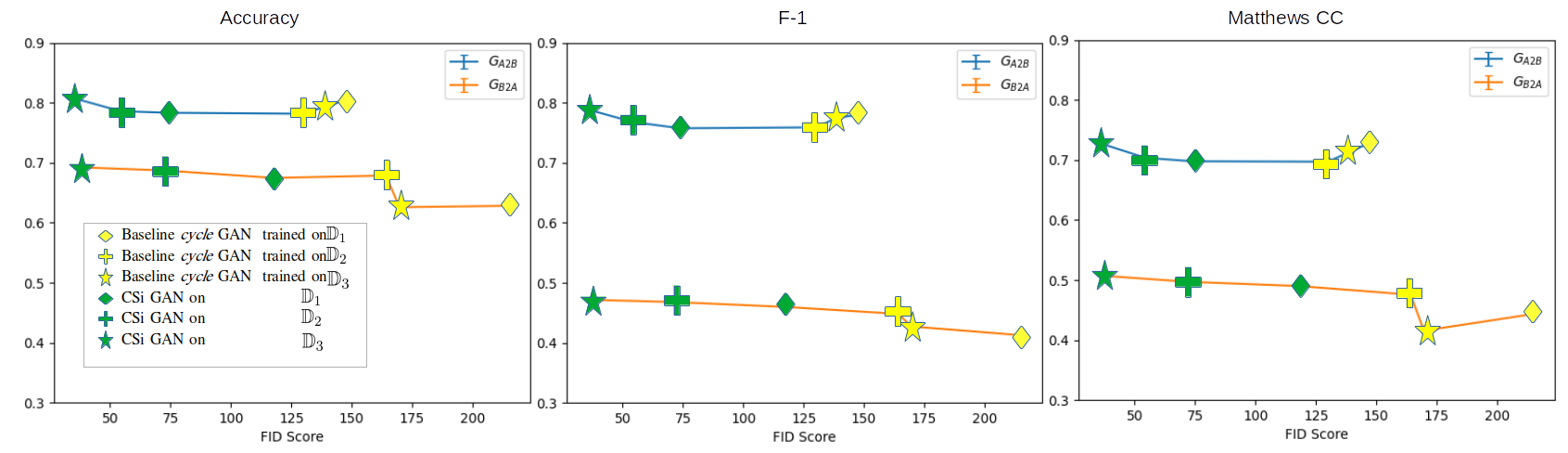}
    \caption{Comparison of the different GAN models when used as backbone for training the multi-input classifier. The results are shown in terms of FID vs :$ACC$, F-1 score, and $MCC$.}
    %\caption{FID comparison of the different GAN models for each of the GAN }
    \label{fig:fid_vs_metrics}
\end{figure*}

An additional analysis was performed in order to determine if the quality of the GAN-translated images influence the classifier performance.  
The metrics $Acc$, F-1 score, and $MCC$, obtained by training the multi-input classifier in a fully supervised using both \emph{Cycle}GAN and CSi-GAN, are compared against the FID score for each of the translation networks. 
The results of this comparison are shown in Fig.~\ref{fig:fid_vs_metrics}. 
Even though it is easy to notice the gap in terms of the FID score between the generators from \emph{Cycle}GAN and  CSi-GAN, and the best classification metrics are obtained when using CSi-GAN with more data~($\mathbb{D}_3$), this improvement is minimum. Furthermore, \emph{Cycle}GAN trained on $\mathbb{D}_2$ obtains similar metrics. 
The comparison against the classification metrics does not show a conclusive result and
further research is needed to determine the correlations that could lead to best practices and parameter choices when training GAN models. 

%The lower the FID value the best quality of the generated images is. 

%It is possible to see that there is indeed a correlation between the amount of data used to train the GAN models and the classification performances, but the major improvement comes from the proposed CSi-GAN, specially for the generator $\mathcal{G}_{BA}$ which translate images from NBI to WLI.  

%It is also importance to notice that the gap in terms of the FID score between the \emph{cycle} generators $\mathcal{G}_{AB}$ and $\mathcal{G}_{BA}$ is reduced when using  CSi-GAN, specially in the case of the models trained on $\mathbb{D}_3$. 

%The results regarding the sensitivity of each of generator model are depicted on Fig.~\ref{fig:noise_analysiss}. 
%Some visual samples of the effects that adding noise to the images has on the generated images are shown on Fig.~\ref{fig:noise_samples}. 
%The models that were trained on a larger amount of data are more resilient to the addition of noise. This is confirmed by the curves depicted on~\ref{fig:noise_analysiss} where the models trained on $\mathbb{D}_2$ and $\mathbb{D}_3$ present lower values as the values of $\sigma$ increases.  This is specially noticeable in the case of the NBI$\rightarrow$WLI translation. The model that presents the best performance in this analysis is again CSi-GAN trained on $D_3$. 
%%%%%%%%%%%

\section{Conclusion}
\label{sec:conclusion}
In this paper, we propose a novel semi-supervised learning GAN-based method to address the problem of endoscopic image classification in NBI and WLI imaging domains.  
The proposed method shows to be effective for a scenario where there is domain and class imbalance and in general, performs better than specialists and baseline methods.
The use of this method leverages the use of unlabeled data in a domain different than the one where annotations exist, which is a very common case in biomedical data where annotated data is limited. 
This could ease the transition to clinical practice and its implementation for computer-aided BC diagnosis. 
The results obtained also show that the quality of the synthetic images generated with the proposed method is good enough to deceive clinical experts. 
Nevertheless, additional research needs to be carried out to find accurate metrics to assess the quality of generated images objectively and to determine to which point it might be related to the classification performances.  

Future work includes further validation of multi-center data, as well as the acquisition of data from other imaging domains which could help to assess better the generalization of the method, and the development of lesion detection methods that could differentiate specific image regions that correspond to the lesion and non-lesion tissue.
By making available our dataset we hope to encourage further research in the field that could motivate the clinical translation of endoscopic image classification. 
%, as well as the development of protocols to  aid in diagnosis and improves surgeons' diagnosis performance.
%
%The developed work 
%investigate methods for 
%Further work also needs to be carried out in order to 
%Validation of this system in the 
%The deployment of this system,  and if this could improves surgeons' diagnosis performance.
%and if this could improves surgeons' diagnosis performance. %Work will also be done in  on multi-center data validation. %that could be used to assess   the use of synthetic 
%whether the use the use of synthetic generated images 

%Limitations: Single center images, bigger datasets, the addition of more domains if available.
%Perspectives: clinical diagnosis, 
%clinical assistance
%
%However, it is important to note that the data used is 
%
%A current issue in biomedical image analysis 
%
%Future work might include multi-center validation. 

\section*{Compliance with ethical standards}
\textbf{Ethical Approval} \\
The proposed study is a retrospective study.
No personal data was recorded.  
The collection of data was in accordance with the ethical standards of the Istituto Europeo di Oncologia and with the 1964 Helsinki declaration, revised in 2000. 
All the subjects involved in this research were informed and agreed to data treatment before the intervention.\\

\textbf{Informed consent}\\
Written informed consent was obtained from all patients included in the study.


%\bibliographystyle{IEEEtran}
%\section*{Acknowledgments}

\bibliographystyle{IEEEtran}
\bibliography{biblio} 
%\begin{thebibliography}{1}
%\end{thebibliography}
%\appendices
%\section*{Appendix I}
%\label{sec:appendix_gradcam}
%\begin{figure*}[tbp]
%    \centering
%    \includegraphics[width=0.5\textwidth]{images/sample_attention_maps.png}
%    \caption{Comparison of the attention maps obtained from the generated images with each of the GAN models and the 3 different datasets. In A)-B) the original domain is WLI, in C)-D) the original domain is NBI. }
%    \label{fig:attention_maps}
%\end{figure*}
%\section*{Appendix II}
%\label{sec:appendix_noise}
%\begin{figure*}
%    \centering
%    \includegraphics[width=0.75\textwidth]{images/noise_analysis.drawio.png}
%    \caption{Curves obtained by comparing the sensitivity ($SN$) of each of the GAN models against the addition of Gaussian noise $N(0, \sigma)$, for different values of $\sigma$. The base-line curve corresponds to the sensitivity calculated between the images without noise and the original image after the addition of that level of $N(0, \sigma)$.}
%    \label{fig:noise_analysiss}
%\end{figure*}

%\begin{figure*}
%    \centering
%    \includegraphics[width=0.99\textwidth]{images/sample_noise.drawio.png}
%    \caption{Image samples of the noise effects for different values of $\sigma$.}
%    \label{fig:noise_samples}
%\end{figure*}
\end{document}